\documentclass[sigconf,nonacm]{acmart}

\setcopyright{none}
\renewcommand\footnotetextcopyrightpermission[1]{}

\usepackage{graphicx}
\usepackage{multirow}
\usepackage{array}
\usepackage{balance}
\usepackage{microtype}
\usepackage{enumitem}
\usepackage[table]{xcolor}
\usepackage{booktabs}

\title{
{\small\textbf{Author copy of the paper accepted for publication in the 30th International Conference on Evaluation and Assessment in Software Engineering (EASE 2026), Glasgow, Scotland.}}\\[1em]
Engineering a Governance-Aware AI Sandbox: Design, Implementation, and Lessons Learned
}

\author{Muhammad Waseem}
\email{muhammad.waseem@tuni.fi}
\affiliation{
  \institution{Faculty of Information Technology and Communication Sciences, Tampere University}
  \city{Tampere}
  \country{Finland}
}

\author{Md Aidul Islam}
\email{mdaidul.islam@tuni.fi}
\affiliation{
  \institution{Faculty of Information Technology and Communication Sciences, Tampere University}
  \city{Tampere}
  \country{Finland}
}

\author{Md Nasir Uddin Shuvo}
\email{nasir.shuvo@tuni.fi}
\affiliation{
  \institution{Faculty of Information Technology and Communication Sciences, Tampere University}
  \city{Tampere}
  \country{Finland}
}

\author{Md Mahade Hasan}
\email{mdmahade.hasan@tuni.fi}
\affiliation{
  \institution{Faculty of Information Technology and Communication Sciences, Tampere University}
  \city{Tampere}
  \country{Finland}
}

\author{Kai-Kristian Kemell}
\email{kai-kristian.kemell@tuni.fi}
\affiliation{
  \institution{Faculty of Information Technology and Communication Sciences, Tampere University}
  \city{Tampere}
  \country{Finland}
}

\author{Jussi Rasku}
\email{jussi.rasku@tuni.fi}
\affiliation{
  \institution{Faculty of Information Technology and Communication Sciences, Tampere University}
  \city{Tampere}
  \country{Finland}
}

\author{Mika Saari}
\email{mika.saari@tuni.fi}
\affiliation{
  \institution{Faculty of Information Technology and Communication Sciences, Tampere University}
  \city{Tampere}
  \country{Finland}
}

\author{Vilma Saari}
\email{vilma.saari@dimecc.com}
\affiliation{
  \institution{DIMECC Ltd}
  \city{Tampere}
  \country{Finland}
}

\author{Roope Pajasmaa}
\email{roope.pajasmaa@dimecc.com}
\affiliation{
  \institution{DIMECC Ltd}
  \city{Tampere}
  \country{Finland}
}

\author{Markku Oivo}
\email{markku.oivo@dimecc.com}
\affiliation{
  \institution{DIMECC Ltd}
  \city{Tampere}
  \country{Finland}
}
\affiliation{
  \institution{University of Oulu}
  \city{Oulu}
  \country{Finland}
}

\author{Pekka Abrahamsson}
\email{pekka.abrahamsson@tuni.fi}
\affiliation{
  \institution{Faculty of Information Technology and Communication Sciences, Tampere University}
  \city{Tampere}
  \country{Finland}
}

\begin{document}

\begin{abstract}
Collaborative AI experimentation in industry--academia requires environments that support rapid trials while maintaining controlled access, organisational isolation, and traceable workflows. Although interest in AI sandboxes is increasing, practical guidance on designing and building governance-aware experimentation platforms remains limited. This work designs and operationalizes a governance-aware, multi-tenant AI sandbox that supports structured experimentation and produces reusable evaluation evidence across stakeholders.

The sandbox was developed in an industry--academia ecosystem using iteratively validated requirements gathered from industrial partners. The solution adopts a layered reference architecture that separates a multi-tenant presentation layer from a backend control plane and isolates execution and data management concerns into dedicated layers. The sandbox supports governed onboarding, project-based collaboration, controlled access to AI services, and traceable experimentation through approval workflows and audit logging. By structuring experiment context and governance decisions as persistent records, the sandbox enables evaluation evidence to be reused and compared across projects and stakeholders.

The development experience yields lessons learned and practical considerations that inform deployment and future evolution of governance-aware sandbox platforms.
\end{abstract}


\maketitle

\section{Introduction}
The adoption of AI technologies, particularly generative AI, is growing across organizations seeking to improve their software systems and business processes, and empirical evidence confirms their increasing integration into software engineering practices \cite{giry2025genaiAdoptionSE}. However, determining the suitability of such technologies for specific organizational contexts remains inherently uncertain, leading some organizations to engage research partners in conducting structured evaluations. While industry–academia collaboration in software engineering has been extensively studied, including its recurring challenges and recommended practices \cite{garousi2016challenges,garousi2019characterizing,marijan2020certus}, how such collaborations unfold specifically in the context of AI technology evaluations remains insufficiently understood.

Despite growing interest, evaluating AI technologies in practice remains difficult. Early trials are often informal and vary across teams, tools, and use cases, which makes results hard to compare and evidence hard to reuse. Industry interview studies report persistent practical issues affecting real-world AI assessment, including governance, compliance, data management, and human oversight \cite{mccormack2025trust}. In addition, research on AI documentation highlights organizational and workflow barriers that limit traceability and reuse of evaluation evidence across stakeholders \cite{winecoff2024aidocumentation}. As a result, feasibility and risk are assessed inconsistently, and decisions about wider adoption are often based on limited evidence. In this paper, evaluation refers to assessing feasibility, risks, and suitability of AI solutions in organizational contexts. To mitigate traceability and reuse challenges described above, we treat approval decisions and audit logs as first-class evaluation artefacts. This aligns with guidance on structured model documentation for comparability \cite{Mitchell2019ModelCards} and with empirical software engineering principles that emphasize explicit artefacts for repeatable evaluation \cite{Wohlin2012Experimentation}.

Existing research has examined regulatory sandboxes and AI governance mechanisms as instruments to balance innovation with oversight, particularly in the context of risk-based regulation and emerging frameworks such as the EU AI Act \cite{TrubyEtAl_SandboxHighRiskAI_2021,BuoczPfotenhauerEisenberger_AIActSandboxes_2023,NovelliEtAl_GettingSandboxesRight_2024}. Related work also positions sandboxes as anticipatory governance tools and collaborative environments that support learning among regulators and stakeholders \cite{Morgan_AnticipatorySandboxSchemes_2023,Macrae_GenerativeSpaces_2024}, and discusses broader experiences of sandboxing for trustworthy AI across regions \cite{Moraes_TrustworthyAI_Sandboxes_2025}. While these studies clarify governance objectives and institutional design considerations, they provide limited guidance on how to architect and operationalize a multi-tenant, governance-aware sandbox infrastructure that supports structured AI experimentation in industry--academia settings \cite{DeGasperis_BridgingTechLaw_2025}. In contrast to work that mainly addresses institutional or regulatory sandbox design, this paper focuses on the technical operationalization of governance-aware experimentation. We contribute a reference architecture and prototype that embed policy enforcement, approval workflows, and structured logging into the sandbox control plane.

\textbf{Motivation}: This work is directly motivated by these limitations and challenges. It emerged from a Finland-based, multi-stakeholder effort within the SW4E ecosystem\footnote{\url{https://sw4e.fi/}}, which aims to make early AI experimentation more trustworthy and reusable across organizations. Tampere University's GPT Lab\footnote{\url{https://gpt-lab.eu/}} led the architecture and prototyping work, while DIMECC\footnote{\url{https://www.dimecc.com/}} coordinated the collaboration to support a vendor-neutral and governed sandbox concept. Industrial stakeholders contributed requirements and feedback through iterative reviews and prototype demonstrations, reinforcing the need for a shared environment that supports controlled AI experimentation and enables evaluation evidence to be captured and reviewed consistently across stakeholders. This ecosystem perspective and the rationale for a shared, trusted environment for collaborative AI experimentation are further documented in the SW4E AI Sandbox white paper \cite{sw4eAISandbox}.


\textbf{Contributions}: This paper makes the following contributions:
\begin{itemize}
    \item We contribute a governance-aware reference architecture and prototype that embed policy enforcement, approval workflows, and structured logging into the sandbox control plane, enabling the operationalization of governed AI experimentation.

    \item An implemented sandbox prototype, released as open source\footnote{\texttt{https://github.com/GPT-Laboratory/GPT-Lab-Sandbox}} and deployed online for demonstration\footnote{\texttt{https://gptlab-frontend-gptlab-sandbox.2.rahtiapp.fi/}}, enabling end-to-end operationalization of governed AI experimentation.

    \item Empirically grounded lessons learned from developing and deploying the sandbox in an industry--academia setting, including a distinction between technical sandboxing and regulatory sandboxing under the EU AI Act.
\end{itemize}

\textbf{Paper Organization}: The remainder of this paper is structured as follows. Section~2 describes the study design and industrial context. Section~3 presents the sandbox design and implementation, including the consolidated requirements, reference architecture, and prototype realisation. Section~4 discusses the findings and reports the lessons learned. Section~5 outlines threats to validity. Section~6 reviews related work. Finally, Section~7 concludes the paper and outlines directions for future work.

\section{Study Design}

\begin{figure*}
    \centering
    \includegraphics[width= 0.8\linewidth]{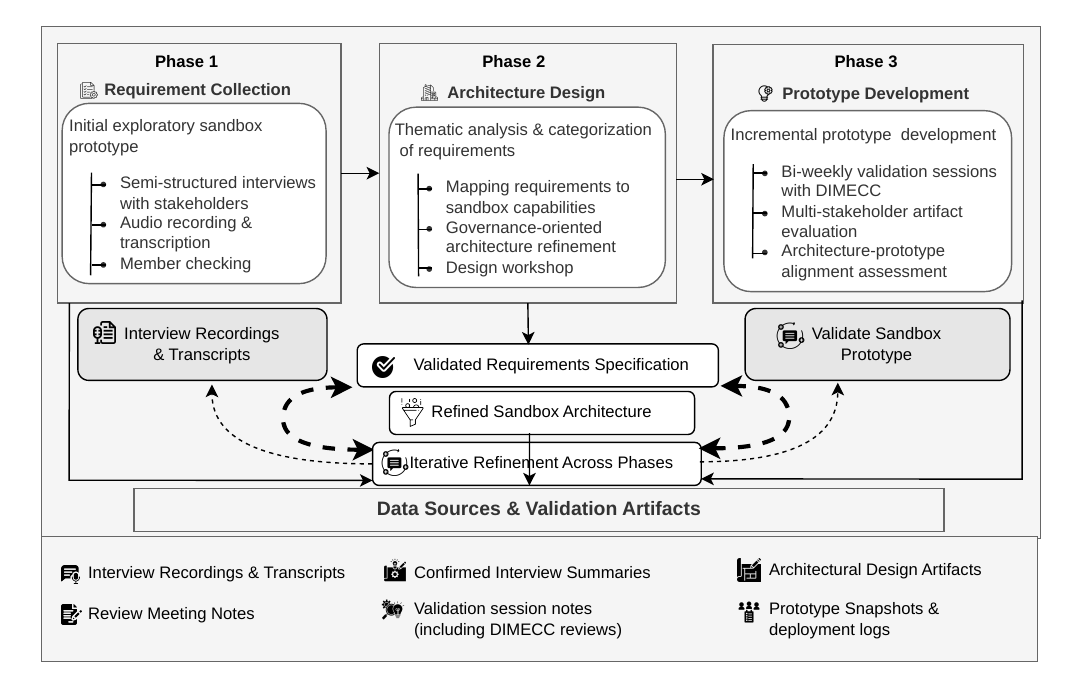}
    \caption{ Overview of the study design}
    \label{fig:Studydesign}
\end{figure*}

The overview of the study is presented in Figure~\ref{fig:Studydesign}. The figure illustrates three iterative phases: requirements collection, architecture design, and prototype development.

\textbf{Context}:
The study was carried out within the Finnish SW4E ecosystem. Tampere University’s GPT Lab led the research and technical development activities, including requirements elicitation, architectural design, governance modelling, and prototype implementation. DIMECC facilitated coordination activities and supported stakeholder engagement. Three industrial partners (Bittium, Q4US, and Solita) participated in the study through structured interviews and iterative review meetings.

\textbf{Phase 1: Requirements Collection}:
The study began with developing an initial sandbox prototype to support structured discussions with industrial stakeholders. Requirements were elicited through semi-structured interviews \cite{seaman1999qualitative} with senior technical representatives from each participating organisation. Each interview lasted approximately one hour and was conducted via Microsoft Teams. Interviews were organised as live walkthrough sessions of the evolving prototype followed by open discussion. Participants were asked to assess existing functionality, identify missing capabilities, discuss governance and security expectations, and suggest improvements. With participants’ consent, sessions were recorded and documented. Recordings were reviewed and transcribed. Requirement statements were extracted from the transcripts and grouped into thematic categories reflecting sandbox functionality, governance mechanisms, security controls, and collaboration workflows. The categorisation was iteratively refined within the research team and shared with the respective interviewees for confirmation to reduce misinterpretation.

\textbf{Phase 2: Architecture Design}:
Validated requirements were translated into a reference architecture for the sandbox environment. Architectural decisions addressed system modularisation, security layering, access control, experiment lifecycle management, and user collaboration support. The architecture evolved incrementally. Updates were introduced whenever new requirements or constraints were identified during interviews or review meetings.

\textbf{Phase 3: Prototype Development}:
A Minimum Viable Prototype (MVP) was incrementally implemented to operationalise the architectural design. Development followed short iterative cycles. Every two weeks, structured prototype review meetings were conducted with DIMECC as the core industrial partner. These sessions focused on evaluating implemented features, identifying limitations, and collecting refinement feedback. Feedback from interviews and review meetings informed subsequent requirement adjustments, architectural modifications, and prototype enhancements. This iterative loop continued throughout the study.

\textbf{Evidence Sources}: As illustrated in Figure~\ref{fig:Studydesign}, the study draws on multiple artefacts produced and validated across phases, including interview recordings and transcripts, confirmed requirement summaries, architectural design artefacts, prototype snapshots, and review meeting notes. These artefacts informed the validated requirements specification and the refined sandbox architecture, both of which evolved iteratively throughout the study.

\section{Sandbox Design and Implementation}

This section presents the design artefacts derived from the industrial case study, including the consolidated requirements, reference architecture, and prototype realisation.

\subsection{Sandbox Requirements}
Using the interview and walkthrough material described in the study context, we extracted stakeholder statements related to functionality and quality attributes and normalised them into candidate requirements. Duplicates were removed, terminology was aligned, and related items were merged. The resulting requirement set was iteratively refined through follow-up discussions and bi-weekly review meetings until agreement was reached on feasibility and scope. The requirements are structured into four main categories.

\textbf{Onboarding and Access Control Requirements}: This category comprises requirements related to user onboarding, access management, and collaboration workflows. Stakeholders required structured onboarding paths tailored to universities and research institutes, companies and organisations, and individual researchers. The sandbox must support self-service registration with email verification, organisation assignment, and role selection, complemented by role-specific entry points. Governed collaboration workflows must enable interactions across stakeholder types (e.g., company--company, university--university, company--university, and individual--organisation). Approval mechanisms should balance usability and governance by combining automated processing for low-risk roles with manual review and escalation for sensitive roles. Role-based dashboards are required to provide immediate, contextualised access to projects and collaboration features after onboarding.
   
\textbf{System and Administrative Requirements}:  This category comprises governance and operational control requirements necessary for managing the sandbox in a multi-tenant environment. The platform must support organisation-level isolation with dedicated policies and dashboards, implemented through multi-layer access control including hierarchical RBAC and least-privilege enforcement. Administrative capabilities must include user lifecycle management (e.g., pending, approved, suspended), approval workflows, and configuration management. To ensure accountability and regulatory alignment, the sandbox must provide comprehensive audit logging of user actions, experiment executions, and data access events. Administrative consoles and system analytics are required to support oversight, monitoring, and operational decision-making.

\textbf{AI Service Requirements}: This category comprises the user-facing AI capabilities delivered through the sandbox under governance constraints. The service catalogue includes planning and decision-support tools (e.g., benchmarking pipelines, cost estimation, and technique assessment), data and experiment lifecycle services (e.g., preprocessing, controlled experiment execution, tracking, and reporting dashboards), and domain-level AI functionalities such as text analytics, computer vision, speech processing, recommendation, time-series analysis, anomaly detection, and LLM experimentation. The sandbox must also support retrieval-augmented generation and model fine-tuning to enable domain adaptation. To ensure responsible and secure operation, AI services must integrate compliance and security mechanisms, including controlled data storage, vulnerability scanning, compliance auditing, and data residency and anonymisation controls. The service layer is designed to be modular and extensible to accommodate future domain-specific additions.
 
 \textbf{Hardware and Infrastructure Requirements}: This category comprises requirements related to provisioning and managing computational resources for AI workloads. The sandbox must provide managed access to high-performance compute (GPU/CPU clusters), scalable storage, and optional cloud-based execution environments for containerised workloads. Resource allocation must be governed through structured request workflows, multi-level approvals, and role-based dashboards displaying allocation status, utilisation, and cost metrics. Infrastructure monitoring and analytics are required to track throughput, utilisation, and per-project consumption to support optimisation and capacity planning. Advanced scheduling mechanisms (e.g., reservations and priority queues) are necessary to ensure fair and efficient resource usage. Security controls, including role-based restrictions, encryption, and audit trails, must protect hardware usage, and integration APIs should enable interoperability with external pipelines and monitoring systems.

\subsection{Architecture Design}
\begin{figure*}
    \centering
    \includegraphics[width=0.9\linewidth]{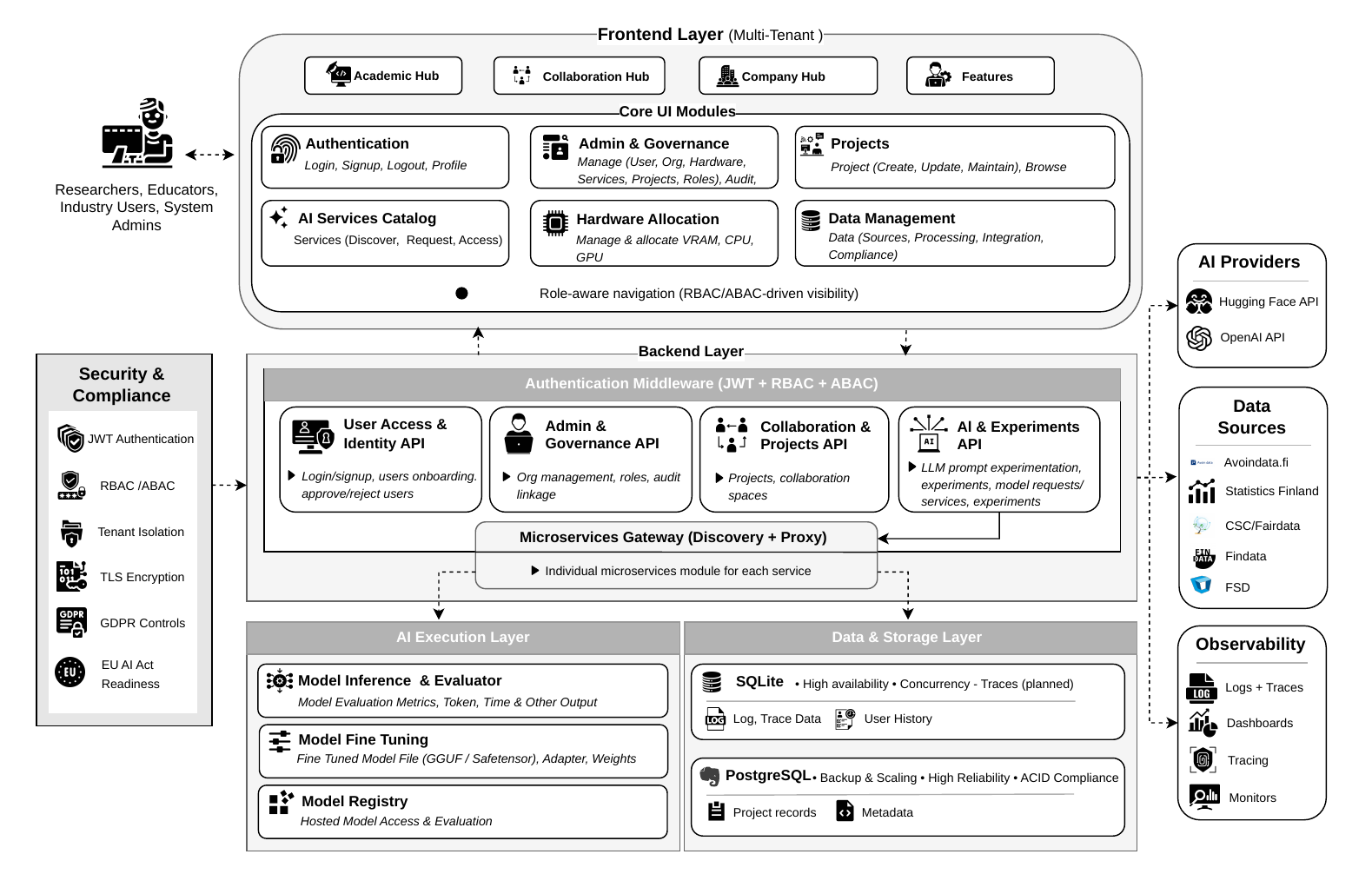}
    \caption{High-level architecture of the AI Sandbox system}
    \label{fig:arch}
\end{figure*}

Figure~\ref{fig:arch} presents the overall architecture of the AI Sandbox for governed experimentation in an industry--academia collaboration setting, highlighting the separation between the governance-focused control plane and the execution layer. The architecture adopts a layered style that enforces clear separation of concerns, explicit policy enforcement, and runtime isolation of AI workload execution from governance functions. The system comprises a multi-tenant presentation layer, a policy-driven control plane mediating access to independently deployed execution services, and persistent data resources. Security, compliance, and observability are treated as cross-cutting mechanisms. 
Due to space constraints, the figure summarises the key components at a high level, while the detailed structure and responsibilities of each layer are described in the following text.

\textit{\textbf{Frontend Layer (Multi-tenant Web UI)}}:
The Frontend Layer realises the user-facing architectural view of the sandbox. It provides three logical workspaces: Academic Workspace, Company Workspace, and Collaboration Workspace, surfaced through dedicated hubs in the UI. These workspaces support distinct stakeholder groups (researchers, educators, industry users, and system administrators) and enable role-specific interaction contexts. The layer comprises core UI modules, including: (i) Authentication, (ii) Admin \& Governance, (iii) Projects, (iv) AI Services Catalogue, (v) Hardware Allocation, and (vi) Data Management. Architecturally, the Frontend Layer operates as a client of the backend control plane and does not directly invoke execution services. Role-aware navigation provides an initial access-filtering mechanism (RBAC/ABAC-driven visibility), reducing invalid interaction paths before backend enforcement. This contributes to defence-in-depth by combining UI-level constraints with server-side policy enforcement.

\textit{\textbf{Backend Layer (Control Plane, REST APIs)}}:  
The Backend Layer realises the control plane of the architecture and encapsulates identity management, governance, collaboration, and experiment coordination. An explicit Authentication Middleware (JWT + RBAC + ABAC) acts as a Policy Enforcement Point (PEP), validating and authorising requests before they are routed to domain services. The control plane is decomposed into four principal APIs aligned with the figure: (i) User Access \& Identity API, supporting login/signup, user onboarding, and approve/reject workflows; (ii) Admin \& Governance API, supporting organisation management, role management, and audit linkage; (iii) Collaboration \& Projects API, supporting projects and collaboration spaces; and (iv) AI \& Experiments API, supporting LLM prompt experimentation and the management of experiments and model/service requests. Requests are routed through a Microservices Gateway (Discovery + Proxy), which proxies calls to the corresponding microservice module for each service. Consistent with the layered design, the Backend Layer coordinates and mediates access to downstream execution and storage layers rather than executing AI workloads directly.

\textit{\textbf{Microservices Gateway (Discovery + Proxy)}}:  
Between the control plane and the execution services, the architecture introduces a Microservices Gateway responsible for service discovery and controlled proxy routing. Architecturally, this gateway acts as a mediation layer, preventing direct coupling between the control-plane APIs and execution microservices. It centralises request routing, enables controlled service exposure, and supports runtime substitution and horizontal scaling of backend services without requiring modifications to the control-plane logic.

\textit{\textbf{AI Execution Layer (Microservices — Deployed Separately)}}:  
The AI Execution Layer represents the execution plane of the architecture. Services in this layer are deployed independently from the control plane to ensure runtime isolation, scalability, and controlled resource utilisation. The layer comprises:  
(i) Model Inference \& Evaluator, responsible for model execution, evaluation metrics, token usage tracking, and time/output measurement;  (ii) Model Fine-Tuning, responsible for managing fine-tuned model artefacts (e.g., GGUF, Safetensors, adapters, and weights); and  
(iii) Model Registry, responsible for hosted model access and evaluation tracking. By separating the control plane from the execution plane, the architecture prevents resource-intensive AI workloads from interfering with governance and coordination logic. This separation supports independent scaling, failure containment, and controlled experimentation.

\textit{\textbf{Data \& Storage Layer}}:  
The Data \& Storage Layer provides persistent storage for operational state, project records, governance metadata, logs, and experiment-related artefacts. The prototype supports SQLite for simplified deployment scenarios and lightweight environments, while PostgreSQL is defined as the production-grade persistence layer. PostgreSQL supports higher concurrency, ACID compliance, structured trace persistence (planned), backup mechanisms, and horizontal scaling capabilities required for enterprise deployment. Logs, user history, and experiment traces are treated as architectural artefacts that provide traceability and accountability across layers.

\textit{\textbf{External Integrations}}: The architecture integrates external AI providers and public data sources through backend-mediated interfaces. These include Hugging Face API and OpenAI API for model access, as well as selected public data sources such as Avoindata.fi, Statistics Finland, CSC/Fairdata, Findata, and FSD. All integrations are accessed through the control plane and gateway mechanisms, ensuring centralised management of credentials, rate limits, and access policies while preventing direct exposure of external endpoints to the frontend or execution layer.

\textit{\textbf{Observability}}:  
Observability is incorporated as a system-wide architectural capability. The architecture includes monitoring dashboards, logs and traces collection, distributed tracing (planned), and OpenShift Observe integration. Observability provides runtime transparency of control-flow paths across layers and supports governance verification, compliance monitoring, and operational diagnostics.

\textit{\textbf{Security and Compliance}}:  
Security and compliance mechanisms span all architectural layers. The architecture enforces JWT-based authentication, combined RBAC and ABAC policies, tenant isolation, TLS encryption, GDPR controls, and EU AI Act readiness considerations. These mechanisms are implemented consistently across the frontend, control plane, execution layer, and data layer, reflecting a cross-cutting architectural concern rather than a localised security component.

\subsection{Implementation}
This section describes the development choices used to realise the proposed architecture, focusing on how governance constraints were enforced in the backend while preserving modular integration and deployment portability.

\textbf{\textit{Frontend Realisation}}: The user-facing layer was implemented as a TypeScript-based Next.js application using the App Router architecture. This design supports modular routing and clearly separated role-specific workspaces. Authorisation is enforced by the backend control plane, while the client relies on backend-validated identity information to enable role-aware navigation and interface-level visibility. Authentication is implemented using JWT-based mechanisms, with token validation performed by backend identity services to eliminate client-side trust assumptions. The frontend technology stack comprises Next.js~14.2, React~18.3, and TypeScript~5.9. This design ensures that governance boundaries remain consistently enforced at the server level.

\textbf{\textit{Backend Control Plane}}: The backend was implemented as a modular monolithic Express application running on Node.js. This approach was chosen to reduce operational overhead during early iterations while preserving clear separation of concerns across authentication, governance enforcement, collaboration workflows, AI service access, and hardware management. Governance rules are enforced through middleware that performs token validation, Role-Based Access Control (RBAC), organisation- and project-level scoping, and permission checks. Domain logic is executed only after successful validation, embedding governance constraints directly into the request lifecycle. The backend technology stack comprises Express~4.19, Node.js~18, and TypeScript~5.4.

\textbf{\textit{Persistence and Audit Strategy}}: SQLite was selected as the prototype datastore to minimise infrastructure dependencies while ensuring transactional consistency. The schema models governance artefacts, including users, roles, organisations, projects, service access rights, hardware allocations, quota records, and audit records. Structured audit logging captures authentication events and governance-relevant actions to support traceability across system layers. The persistence layer is implemented using \texttt{better-sqlite3} with write-ahead logging enabled. The current configuration prioritises architectural validation and functional correctness over production-scale optimisation.

\textbf{\textit{AI Integration Approach}}: AI capabilities are integrated through backend-managed adapters that invoke external inference APIs via synchronous HTTP requests. Credentials and rate limits are handled exclusively on the server side to prevent exposure to client environments. AI services are encapsulated as modular components behind stable control-plane interfaces, enabling incremental extension without modifying governance enforcement mechanisms. A standalone Python-based anonymisation microservice illustrates the extensibility of the integration model. The integration layer currently supports providers such as Hugging Face and OpenAI, with credentials supplied through environment variables. 
To mitigate privacy risks when interacting with external AI providers, the sandbox enforces server-side credential management and restricts direct client-side access to external APIs. Sensitive data can be processed through anonymisation mechanisms before being transmitted to external services, as illustrated by the integration of a dedicated anonymisation microservice. In addition, all external interactions are mediated through the control plane, ensuring that data access is governed, logged, and subject to policy enforcement. In a fuller production-oriented deployment, these controls are intended to be complemented by stronger data protection measures such as stricter encryption and access policies, clearer data residency constraints, and enhanced organisational isolation for multi-party collaboration.

\textbf{\textit{Hardware Governance Implementation}}: Hardware allocation is implemented through governed workflows rather than direct cluster integration. GPU, CPU, memory, storage, and network resources are represented as logical entities in the data model, and allocation decisions are mediated through approval workflows and quota tracking. The prototype simulates compute resources, prioritising governance validation over infrastructure orchestration. The implementation includes data structures for resource requests, approvals, and allocations, governed by role-based administrative controls.

Table~\ref{tab:implementation-stack} summarises the technology stack used in the implementation.

\begin{table}[htbp]
\centering
\scriptsize
\caption{Technology stack used in the implementation of the AI Sandbox}
\label{tab:implementation-stack}
\begin{tabular}{p{2.2cm} p{5.8cm}}
\toprule
\rowcolor{gray!15}
\textbf{Module} & \textbf{Technologies and Libraries Used} \\
\midrule
Frontend & Next.js~14 (App Router), React~18, TypeScript~5, Tailwind CSS, Radix UI, TanStack React Query, JWT-based authentication (Bearer token; optional HTTP-only cookie in deployment variant). \\
Backend & Express~4 on Node.js~18, TypeScript, REST APIs, middleware-based RBAC and organisation/project scoping, subscription validation. \\
Database & SQLite with \texttt{better-sqlite3}, raw SQL queries, WAL mode enabled, foreign-key constraints enforced. \\
AI Integration & External inference APIs (Hugging Face, OpenAI) invoked via synchronous HTTP calls; environment-based credential management; server-side rate limiting. \\
Domain Microservice & Python FastAPI-based data anonymisation service using Microsoft Presidio and spaCy; independently containerised. \\
Containerisation & Multi-stage Docker builds, Docker Compose for local orchestration, Kubernetes-compatible deployment. \\
\bottomrule
\end{tabular}
\end{table}

\subsection{Testing}

This subsection summarises the verification activities conducted to validate the sandbox prototype. Testing focused on workflow correctness and governance enforcement rather than large-scale performance evaluation.

\textbf{\textit{Functional Verification}}: Core workflows, including user registration, approval handling, role-based dashboard access, project creation, AI service invocation, and hardware request submission—were executed manually in a structured manner. Particular attention was given to verifying that governance constraints (role checks, organisation and project scoping, and permission validation) were consistently enforced at API boundaries. Health endpoints were used to confirm service availability in both local and containerised environments.

\textbf{\textit{Access-Control and Governance Testing}}: Access-control enforcement was evaluated by attempting unauthorised operations across role and organisation boundaries. These included restricted administrative actions, cross-organisation resource access attempts, and service invocation without required permissions. Middleware-based RBAC and permission checks correctly rejected invalid requests. Audit logs were reviewed to confirm that security-relevant events were properly recorded.

\textbf{\textit{Automation and Coverage}}: The data anonymisation microservice includes a single frontend test file (\texttt{App.test.tsx}), which serves as a default placeholder and does not exercise anonymisation logic. No comprehensive automated test suite has been implemented for the main frontend or backend components. A lightweight CI configuration validates build consistency (control plane and UI) and container integrity for template images (Trivy scanning and Cosign signing). Testing activities prioritised architectural validation and governance enforcement over full automation and performance benchmarking.

\subsection{Deployment}
 The system is containerised and deployable on Kubernetes-compatible platforms (e.g., OpenShift) to validate runtime portability and operational feasibility. Core components are packaged as Docker containers and can be orchestrated locally using Docker Compose or deployed to a managed Kubernetes environment. Deployment updates are performed manually, with service health monitored using platform-level mechanisms.  Runtime configuration and external service credentials are supplied via environment variables. No dedicated secret-management integration is implemented at this stage. The persistence layer currently relies on SQLite; migration to a production-grade relational database (e.g., PostgreSQL) is anticipated for concurrent multi-user workloads. The deployment configuration prioritises architectural and governance validation over production-scale automation.
\section{Discussion and Lessons Learned}

This study progressed from elicited requirements to a realised prototype in order to understand what is required to construct a governance-aware AI sandbox for industry--academia collaboration. Beyond the architectural artefacts presented earlier, the development process exposed structural, operational, and regulatory considerations that extend beyond technical implementation.

From a practical perspective, the sandbox provides value by enabling organisations to structure early-stage AI experimentation under governance constraints. Rather than relying on informal and isolated trials, the platform supports traceable workflows, reusable evaluation artefacts, and consistent access control, allowing stakeholders to compare results across projects and make more informed adoption decisions. While the sandbox does not replace formal regulatory oversight, it improves transparency and accountability within organisational boundaries. This also highlights the distinction between technical mechanisms for governed experimentation and broader regulatory sandboxing frameworks, which is discussed next.

\subsection{Discussion}

\textit{\textbf{Technical sandboxing versus regulatory sandboxing}}: The implemented prototype demonstrates technical mechanisms required for controlled AI experimentation, including role-based access control (RBAC), organisation and project scoping, approval workflows, and structured audit logging. However, these mechanisms alone do not constitute an Article~53-style regulatory sandbox under the EU AI Act. Prior research emphasises that regulatory sandboxes require competent authority supervision, formally agreed sandbox plans, and institutional accountability mechanisms beyond software infrastructure \cite{BuoczPfotenhauerEisenberger_AIActSandboxes_2023,NovelliEtAl_GettingSandboxesRight_2024,Morgan_AnticipatorySandboxSchemes_2023}.The prototype should therefore be understood as a governance-aware experimentation environment rather than a regulatory sandbox. This distinction prevents overstating compliance while clarifying how technical infrastructure can support supervisory processes.

\textit{\textbf{Governance as an architectural driver}}: Governance requirements directly shaped API boundaries, workflow design, and persistence structures. The need for traceability and auditable decision paths motivated an explicit control plane in which authentication, authorisation, approval logic, and logging are enforced prior to service execution. This aligns with broader ``by-design'' governance principles, which argue that regulatory constraints must be structurally embedded rather than retrofitted \cite{Cavoukian_PrivacyByDesign_2010,TrubyEtAl_SandboxHighRiskAI_2021}. The study operationalises this perspective at the architectural level by modelling audit evidence and policy enforcement as first-class system concerns.

\textit{\textbf{Iterative validation before infrastructure optimisation}}: Several implementation decisions, including lightweight persistence (SQLite), manual deployment workflows, simulated hardware allocation (governed request--approval workflows without scheduler integration), and limited test automation reflect a deliberate emphasis on validating governance logic and collaboration workflows before pursuing production-scale optimisation. This approach mirrors the conception of regulatory sandboxes as experimental governance instruments where learning precedes institutional stabilisation \cite{Morgan_AnticipatorySandboxSchemes_2023}. The experience suggests that sandbox initiatives benefit from stabilising control-plane assumptions before investing in orchestration complexity and performance engineering.

\textit{\textbf{Deployment environments shape system design}}: Platform-level constraints, including container policies and runtime permission models, influenced architectural decisions and required early validation. Governance-aware systems therefore benefit from deployment-aligned validation rather than purely local prototyping. This observation is consistent with research emphasising that sandbox environments are shaped by institutional and infrastructural contexts \cite{Macrae_GenerativeSpaces_2024}.

\textit{\textbf{Socio-technical dependencies of adoption}}: Although governance enforcement can be embedded in software, operational readiness depends on organisational and legal arrangements such as liability boundaries, intellectual property constraints, approval hierarchies, and regulatory interpretation. Research on trustworthy AI governance highlights that compliance emerges from socio-technical configurations rather than technical artefacts alone \cite{Moraes_TrustworthyAI_Sandboxes_2025}. 

\subsection{Lessons Learned}

The iterative development and validation of the sandbox prototype resulted in the following consolidated lessons, reflecting architectural, operational, and regulatory considerations observed during stakeholder engagement and implementation.

\begin{enumerate}
    \item \textbf{\textit{Governance must be structural, not additive}}:  
    Access control, organisation and project scoping, approval workflows, and audit logging are most effective when embedded within the control plane middleware rather than introduced as feature-level extensions. Composable middleware (e.g., \texttt{requireRole}, \texttt{requireOrganizationAccess}) enables fine-grained enforcement but requires consistent application across endpoints to avoid gaps.

    \item \textbf{\textit{Technical sandboxing and regulatory sandboxing are distinct constructs}}:  
    A governance-aware experimentation platform can support innovation, but formal regulatory sandboxing under the EU AI Act Article~53 requires competent authority supervision, sandbox plans, and institutional accountability mechanisms beyond software architecture.

    \item \textbf{\textit{Early architectural decisions shape compliance scalability}}:  
    Persistence models (e.g., SQLite versus PostgreSQL), logging strategies, authentication mechanisms, and role-permission structures directly influence the feasibility of later production-grade or compliance-oriented deployment. Prototype choices such as lightweight persistence and file-level locking proved adequate for validation but revealed concurrency limits under concurrent writes.

    \item \textbf{\textit{Environmental consistency reduces governance ambiguity}}:  
    Alignment between development and deployment configurations strengthens enforcement clarity and reduces security inconsistencies. In this prototype, dual frontend variants and divergent audit implementations (structured logging in deployment versus console logging in development) created drift that required explicit reconciliation.

    \item \textbf{\textit{Deployment constraints influence architectural structure}}:  
    Container policies, runtime permissions, and infrastructure boundaries actively shape implementation decisions and should be validated early. Platform-specific constraints, such as OpenShift dynamic user IDs requiring explicit writable storage paths emerged only at deployment and necessitated architectural adjustments.

    \item \textbf{\textit{Simulated governance validates workflow logic, not real enforcement}}:  
    Approval flows and quota models support conceptual evaluation of governance design but do not substitute for integration with real infrastructure (e.g., schedulers, resource allocators) or supervisory authority processes. Hardware allocation in this prototype used governed request--approval workflows with logical resource entities but no actual cluster binding.

    \item \textbf{\textit{Auditability influences stakeholder trust more than AI capability depth}}:  
    Observations during development and validation suggested that traceability, approval transparency, and audit evidence were prioritised over advanced model functionality when assessing sandbox readiness. Structured audit logging and governance visibility proved central to stakeholder confidence.

    \item \textbf{\textit{Regulatory readiness is inherently socio-technical}}:  
    Alignment with the EU AI Act depends on coordination between architecture, organisational processes, legal interpretation, and competent authorities. The prototype's technical platform achieved partial alignment (e.g., RBAC, audit trails, GDPR features), but institutional elements, such as authority designation and sandbox plan formalisation, remain outside the scope of software implementation alone.
\end{enumerate}

\section{Threats to Validity}

This study is subject to several limitations that may influence the interpretation of its results and the scope of its applicability.

 \textbf{Construct validity}:  
  The study focuses on architectural feasibility, governance enforcement mechanisms, and workflow behaviour rather than quantitative performance, usability evaluation, or large-scale adoption outcomes.  Consequently, the reported insights are grounded in structural analysis, prototype realisation, and stakeholder validation rather than controlled benchmarking or empirical user studies. Quantitative evaluation of performance, adoption effort, and governance overhead is outside the scope of this study and remains an important direction for future work. Defining governance-oriented evaluation metrics and exploring automated assessment approaches, such as LLM-based evaluation frameworks, represent promising directions for future research.
  

\textbf{Internal validity}:  
  The prototype was developed iteratively, and several implementation decisions, such as simulated hardware governance, manual deployment workflows, and limited automated testing, were deliberate trade-offs to prioritise architectural and governance validation. These decisions may affect operational robustness and limit conclusions regarding performance scalability or production readiness.

 \textbf{External validity}:  
  The sandbox was designed and evaluated within a specific Finnish industry--academia ecosystem. Governance practices, compliance expectations, infrastructural constraints, and collaboration models may differ across sectors or jurisdictions. As such, generalisation to other organisational or regulatory contexts should be undertaken with caution.

 \textbf{Regulatory validity}:  
  The alignment discussion with the EU AI Act reflects a technical interpretation of Article~53-type regulatory sandbox requirements based on architectural and workflow characteristics. Formal regulatory-sandbox status depends on competent authority supervision and institutional processes beyond the implemented system. Moreover, evolving guidance and national implementation practices may affect how regulatory provisions are interpreted in practice.





\section{Related Work}

Research relevant to AI sandbox environments can be grouped into two main streams. The first stream discusses regulatory sandboxes as governance instruments designed to support innovation while enabling supervision and institutional learning. The second stream focuses on AI governance and trustworthy AI in relation to risk classification, compliance, and operational accountability requirements. However, existing work rarely connects these perspectives to concrete system-level design for multi-tenant AI sandbox infrastructures.

\subsection{AI Governance and Risk-Based Regulation}

A body of work addresses the broader governance landscape for artificial intelligence systems and emphasizes that emerging regulatory frameworks increasingly adopt a risk-based logic. Within this framing, scholars highlight the tension between innovation and safety, particularly for high-risk AI applications, and propose structured experimentation environments as a mechanism to balance these objectives. Regulatory sandboxes are frequently discussed as instruments that allow controlled experimentation under supervisory oversight while reducing regulatory uncertainty.

For instance, prior work on high-risk AI applications argues that sandbox mechanisms can complement strict regulatory regimes by enabling supervised evaluation before large-scale deployment \cite{TrubyEtAl_SandboxHighRiskAI_2021}. Similarly, studies examining regulatory sandboxes under the European Union AI Act emphasize their potential to reconcile innovation incentives with safety, accountability, and compliance expectations \cite{BuoczPfotenhauerEisenberger_AIActSandboxes_2023}. Additional research positions regulatory sandboxes as anticipatory governance tools that allow regulators and stakeholders to iteratively refine oversight practices in response to technological evolution \cite{Morgan_AnticipatorySandboxSchemes_2023}.

While these studies provide important conceptual foundations for understanding the role of sandboxing in AI governance, their emphasis remains primarily on policy objectives, institutional arrangements, and legal considerations. 

\subsection{Regulatory Sandbox Design under the AI Act}

A second stream of research focuses more directly on the design and governance of regulatory sandboxes within the context of the EU AI Act and related regulatory frameworks. These studies analyze how sandbox mechanisms are structured, the roles of supervisory authorities, and the procedural requirements for participation.

For example, work on the governance of regulatory sandboxes under the AI Act discusses institutional design choices, lifecycle phases, and oversight mechanisms that shape sandbox effectiveness \cite{NovelliEtAl_GettingSandboxesRight_2024}. Related research examines the interaction between AI regulatory sandboxes and data protection law, particularly the General Data Protection Regulation, and highlights the need for coherent compliance approaches across regulatory domains \cite{BaldiniFrancis_AIActGDPR_2024}. Other studies extend the analysis to cybersecurity considerations, emphasizing that sandbox frameworks must integrate security-by-design principles to address emerging risks \cite{Bagni_CybersecurityChallenge_2023}.

Broader perspectives also explore global and regional implementations of regulatory sandboxes, including comparative analyses and experiences from Latin American contexts, framing sandboxes as instruments to promote trustworthy AI development \cite{Moraes_TrustworthyAI_Sandboxes_2025}. Furthermore, research highlighting the need to bridge technological implementation and legal oversight underscores the socio-technical complexity inherent in operationalizing regulatory sandbox environments \cite{DeGasperis_BridgingTechLaw_2025}.

Despite providing valuable insights into governance structures and regulatory intent, these studies typically remain conceptual or legal in orientation. They do not provide detailed architectural blueprints, multi-tenant isolation models, workflow designs, or technical enforcement mechanisms required to implement an operational AI sandbox platform.

\subsection{Toward Integrated Sandbox Architectures}

Recent work increasingly acknowledges that sandboxing involves more than isolated testing environments and must support collaboration, learning, and traceable evaluation processes \cite{Macrae_GenerativeSpaces_2024}. This perspective reinforces the view that AI sandboxes are socio-technical infrastructures requiring both governance alignment and technical realization.

Overall, the reviewed literature demonstrates strong interest in regulatory sandboxing as a governance mechanism and in AI risk management frameworks. However, there is limited research that integrates regulatory objectives, security requirements, operational workflows, and system-level architectural design into a unified sandbox reference model. In contrast to prior work, this study proposes a requirements-driven and governance-aligned sandbox architecture, validated through industry collaboration and prototype implementation, that structurally embeds compliance, supervision, and secure experimentation into a coherent multi-tenant AI infrastructure.
\section{Conclusion}

This study presented the design and implementation of a governance-aware AI sandbox to support structured collaboration between industry--academia. The architecture separates user-facing functionality from backend governance and isolates execution and data management into dedicated layers. The implemented sandbox supports controlled onboarding, role-based access, project collaboration, approval workflows, and audit logging to enable traceable AI experimentation. Requirements were elicited and iteratively validated with industrial stakeholders, and the paper reported lessons learned on governance by design, deployment constraints, and socio technical alignment. The study also clarified the difference between technical sandbox infrastructure and broader regulatory arrangements that depend on institutional supervision and processes beyond software implementation. Future work will focus on improving automation and audit completeness, strengthening resource governance, integrating production-grade persistence, and evaluating the platform in larger multi-organisation deployments.

\section*{Acknowledgments}

The authors used ChatGPT (OpenAI) only for editorial assistance (grammar, style, and clarity) and not for generating research content. The authors reviewed and approved the final manuscript.

\bibliographystyle{ACM-Reference-Format}
\bibliography{acm-ref}

@misc{giry2025genaiAdoptionSE,
  title         = {An Empirical Study of Generative AI Adoption in Software Engineering},
  author        = {Giray, G{\"o}rkem and Demir{\"o}rs, Onur and Kalinowski, Marcos and Mendez, Daniel},
  year          = {2025},
  month         = dec,
  eprint        = {2512.23327},
  archivePrefix = {arXiv},
  primaryClass  = {cs.SE},
  url           = {https://arxiv.org/abs/2512.23327}
}

@article{garousi2016challenges,
  title   = {Challenges and Best Practices in Industry--Academia Collaborations in Software Engineering: A Systematic Literature Review},
  author  = {Garousi, Vahid and Petersen, Kai and Ozkan, Baris},
  journal = {Information and Software Technology},
  volume  = {79},
  pages   = {106--127},
  year    = {2016},
  doi     = {10.1016/j.infsof.2016.07.006},
  url     = {https://www.sciencedirect.com/science/article/pii/S0950584916301203}
}

@article{garousi2019characterizing,
  title   = {Characterizing Industry--Academia Collaborations in Software Engineering: Evidence from 101 Projects},
  author  = {Garousi, Vahid and Felderer, Michael and Fernandes, Jo{\~a}o M.},
  journal = {Empirical Software Engineering},
  year    = {2019},
  doi     = {10.1007/s10664-019-09711-y},
  url     = {https://link.springer.com/article/10.1007/s10664-019-09711-y}
}

@article{marijan2020certus,
  title   = {Industry-Academia Research Collaboration in Software Engineering: The Certus Model},
  author  = {Marijan, Dusica and Gotlieb, Arnaud},
  journal = {Information and Software Technology},
  year    = {2020},
  doi     = {10.1016/j.infsof.2020.106473},
  url     = {https://www.sciencedirect.com/science/article/pii/S0950584920302184}
}

@article{mccormack2025trust,
  title   = {Trust and transparency in {AI}: industry voices on data, ethics, and compliance},
  author  = {McCormack, Louise and Bendechache, Malika and Lewis, Dave and Huyskes, Diletta},
  journal = {AI \& Society},
  year    = {2025},
  doi     = {10.1007/s00146-025-02654-7},
  url     = {https://link.springer.com/article/10.1007/s00146-025-02654-7},
  note    = {Published 15 Oct 2025}
}

@misc{winecoff2024aidocumentation,
  title         = {Improving governance outcomes through {AI} documentation: Bridging theory and practice},
  author        = {Winecoff, Amy A. and Bogen, Miranda},
  year          = {2024},
  eprint        = {2409.08960},
  archivePrefix = {arXiv},
  primaryClass  = {cs.HC},
  doi           = {10.48550/arXiv.2409.08960},
  url           = {https://arxiv.org/abs/2409.08960},
  note          = {arXiv v2, 9 Dec 2024}
}

@misc{sw4eAISandbox,
  title        = {AI Sandbox: Secure, Transparent and Collaborative AI Experimentation},
  author       = {{SW4E Ecosystem}},
  institution  = {DIMECC Oy},
  year         = {2025},
  address      = {Finland},
  note         = {Industry white paper},
  url          = {https://www.dimecc.com/wp-content/uploads/2025/11/AI-Sandbox.pdf},
}

@article{seaman1999qualitative,
  author    = {Carolyn B. Seaman},
  title     = {Qualitative Methods in Empirical Studies of Software Engineering},
  journal   = {IEEE Transactions on Software Engineering},
  volume    = {25},
  number    = {4},
  pages     = {557--572},
  year      = {1999},
  publisher = {IEEE}
}

@article{TrubyEtAl_SandboxHighRiskAI_2021,
  title={A sandbox approach to regulating high-risk artificial intelligence applications},
  author={Truby, Jon and Brown, Rafael Dean and Ibrahim, Imad Antoine and Parellada, Oriol Caudevilla},
  journal={European Journal of Risk Regulation},
  volume={13},
  number={2},
  pages={270--294},
  year={2022},
  publisher={Cambridge University Press}
}

@article{BuoczPfotenhauerEisenberger_AIActSandboxes_2023,
  title={Regulatory sandboxes in the AI Act: reconciling innovation and safety?},
  author={Buocz, Thomas and Pfotenhauer, Sebastian and Eisenberger, Iris},
  journal={Law, Innovation and Technology},
  volume={15},
  number={2},
  pages={357--389},
  year={2023},
  publisher={Taylor \& Francis}
}

@inproceedings{Morgan_AnticipatorySandboxSchemes_2023,
  title={Anticipatory regulatory instruments for AI systems: A comparative study of regulatory sandbox schemes},
  author={Morgan, Deborah},
  booktitle={Proceedings of the 2023 AAAI/ACM Conference on AI, Ethics, and Society},
  pages={980--981},
  year={2023}
}

@article{NovelliEtAl_GettingSandboxesRight_2024,
  title={Getting regulatory sandboxes right: Design and governance under the AI Act},
  author={Novelli, Claudio and Hacker, Philipp and McDougall, Simon and Morley, Jessica and Rotolo, Antonino and Floridi, Luciano},
  journal={Available at SSRN 5332161},
  year={2025}
}

@inproceedings{BaldiniFrancis_AIActGDPR_2024,
  title={AI Regulatory Sandboxes between the AI Act and the GDPR: the role of Data Protection as a Corporate Social Responsibility},
  author={Baldini, Davide and Francis, Kate},
  booktitle={CEUR Workshop Proceedings},
  volume={3731},
  year={2024},
  organization={CEUR-WS}
}

@article{Bagni_CybersecurityChallenge_2023,
  title={The regulatory sandbox and the cybersecurity challenge: from the artificial intelligence act to the cyber resilience act},
  author={Bagni, Filippo},
  journal={Rivista italiana di informatica e diritto},
  volume={5},
  number={2},
  pages={201--217},
  year={2023}
}

@article{Moraes_TrustworthyAI_Sandboxes_2025,
  title={Regulatory sandboxes for trustworthy artificial intelligence--global and Latin American experiences},
  author={Moraes, Thiago},
  journal={International Review of Law, Computers \& Technology},
  volume={39},
  number={1},
  pages={55--74},
  year={2025},
  publisher={Taylor \& Francis}
}

@article{DeGasperis_BridgingTechLaw_2025,
  title={Regulatory Sandboxes in Artificial Intelligence: bridging the gap between technology and law},
  author={De Gasperis, Ilaria},
  journal={Artificial Intelligence},
  volume={3},
  number={3},
  year={2025}
}

@article{Macrae_GenerativeSpaces_2024,
  author  = {Macrae, Carl and Ansell, Christopher K.},
  title   = {Generative Spaces: Collaboration, Learning and Innovation in a Regulatory Sandbox},
  journal = {SSRN Electronic Journal},
  year    = {2024},
  month   = may,
  note    = {Available at SSRN: \url{https://ssrn.com/abstract=4825907}},
  doi     = {10.2139/ssrn.4825907}
}

@misc{Cavoukian_PrivacyByDesign_2010,
  author       = {Ann Cavoukian},
  title        = {Privacy by Design: The 7 Foundational Principles},
  year         = {2010},
  institution  = {Information and Privacy Commissioner of Ontario},
  url          = {https://student.cs.uwaterloo.ca/~cs492/papers/7foundationalprinciples_longer.pdf}
}

@inproceedings{Mitchell2019ModelCards,
  title={Model cards for model reporting},
  author={Mitchell, Margaret and Wu, Simone and Zaldivar, Andrew and Barnes, Parker and Vasserman, Lucy and Hutchinson, Ben and Spitzer, Elena and Raji, Inioluwa Deborah and Gebru, Timnit},
  booktitle={Proceedings of the conference on fairness, accountability, and transparency},
  pages={220--229},
  year={2019}
}

@book{wohlin2012experimentation,
  title={Experimentation in software engineering},
  author={Wohlin, Claes and Runeson, Per and H{\"o}st, Martin and Ohlsson, Magnus C and Regnell, Bj{\"o}rn and Wessl{\'e}n, Anders and others},
  volume={236},
  year={2012},
  publisher={Springer}
}

\balance
\end{document}